\documentclass[showpacs,preprintnumbers,amsmath,amssymb,superscriptaddress]{revtex4}
\usepackage{amsmath}
\usepackage{amssymb}
\usepackage{amsfonts}
\usepackage{fancyhdr}
\usepackage{fancybox}
\usepackage{ulem}
\usepackage{color}
\usepackage{graphicx}
\usepackage{float}

\usepackage{amsthm}

\DeclareMathOperator{\Tr}{Tr}

\makeatletter
\def\loweq@align#1#2{\lower.6ex\vbox{\baselineskip\z@skip\lineskip\z@
    \ialign{$\m@th#1\hfil##\hfil$\crcr#2\crcr=\crcr}}}
\def\lowsim@align#1#2{\lower.6ex\vbox{\baselineskip\z@skip\lineskip\z@
    \ialign{$\m@th#1\hfil##\hfil$\crcr#2\crcr\sim\crcr}}}
\def\geqq{\mathrel{\mathpalette\loweq@align >}}
\def\leqq{\mathrel{\mathpalette\loweq@align <}}
\def\grsim{\mathrel{\mathpalette\lowsim@align >}}
\def\lesssim{\mathrel{\mathpalette\lowsim@align <}}
\def\gsim{\mathrel{\mathpalette\lowsim@align >}}
\def\lsim{\mathrel{\mathpalette\lowsim@align <}}
\makeatother
\newcommand{\grless} 
{ {\, \raise-.24em\hbox{$<$} \hspace{-0.8em} \raise.31em\hbox{$>$}\, } }
\newcommand{\lessgr} 
{ {\, \raise-.24em\hbox{$>$} \hspace{-0.8em} \raise.31em\hbox{$<$}\, } }
\newfont{\bg}{cmr10 scaled\magstep4}                    
\newcommand{\bigzerou}{\smash{\lower1.7ex\hbox{\bg 0}}}

\newcommand{\crl}[1]{[-\infty,\infty]}

\newcommand{\ket}[1]{|{#1}\rangle}

\newcommand{\bra}[1]{\langle{#1}|}
\newcommand{\bkt}[2]{\langle{#1}|{#2}\rangle}

\usepackage{amsthm}

\newtheorem*{theo}{Theorem}

\begin{document}
\title{Weak Values as Context Dependent Values of Observables and Born's Rule}
\author{Akio Hosoya}\email{ahosoya@th.phys.titech.ac.jp}
 \author{Minoru Koga}\email{koga@th.phys.titech.ac.jp} 
 \affiliation{Department of Physics, Tokyo Institute of
 Technology, Tokyo, Japan}
\date{\today}

\begin{abstract}
We characterize a value of an observable by a `sum rule' for generally non-commuting observables and a `product rule'
when restricted to a maximal commuting subalgebra of observables together with 
the requirement that the value is unity for the projection operator of the prepared state
and the values are zero for the projection operators of the states which are orthogonal
 to the prepared state. 
The crucial requirement is that  the expectation value and the variance of an observable should be independent of the way of measurement,
 i.e., the choice of the maximal commuting subalgebra of observables.
 We shall call the value a {\it `contextual value'}.
 We show that the contextual  value of an observable coincides with the weak value advocated
by Aharonov and his colleagues by demanding the consistency of quantum mechanics
with  Kolmogorov's measure theory of probability.
 This also gives a derivation of Born's rule, which is one
 of the axioms of conventional quantum mechanics.
 \end{abstract}


\maketitle

\section{Introduction}
 In the conventional Copenhagen interpretation of quantum mechanics, the value of physical quantity emerges
 only after measurement but not before. 
However, the idea of value only after measurement becomes  problematic when one considers quantum gravity, 
 because measurements presuppose spacetime which emerges after the quantum era of the Universe apart from the fact that
 there are no observers who can make projective measurements of the Universe ~\cite{ID}. Independently 
 Ozawa~\cite{Ozawa} has pursued the possibility of assigning a context dependent value for a physical quantity before measurement,
 where the context  describes outcomes of the measurements to be performed.
  
   On the other hand, Born's rule~\cite{Born} is an axiom of conventional quantum mechanics, 
   though there is literature which
   claims that the rule can be derived. It seems that there is a lack of either physical intuition because of a lack of
   a measurement process or a lack of mathematical rigor.
   We shall give a brief review of these notions restricting our attention to the orthodox view in the discussion part of the present work but not to the many world interpretation of quantum mechanics~\cite{many,deutsch}.
 
      In the present paper we show that the contextual value of an observable coincides with the weak value proposed
by Aharonov and his colleagues ~\cite{AAV,AR} on the basis of consistency of quantum mechanics
with  Kolmogorov's measure theory of probability ~\cite{Kolmo} and that this consistency also leads to Born's rule as the unique probability
measure of events to be discovered by the post-selection.

 The organization of the present paper is as follows. After a brief preliminary in Sect 2, we present in Sect 3 a theorem
that the weak values are the contextual values of observables and give a derivation of Born's rule. 
The final section is a summary and discussion of our results and includes a brief review of the claims of the derivation of Born's rule.
    
\section{Preliminary}
    
 We first set up the general framework. Let $\cal{N}$ be a set of observables which act on the Hilbert space 
    $\cal{H}$ of finite dimension, i.e., dim${\mathcal H}=N$. 
   {Let $\mathcal{V(N)}_{\rm max}$ be the set of all maximal abelian subalgebras of $\mathcal{N}$ and 
   choose a maximal abelian subalgebra $V_{\rm max}\in{\cal V(N)}_{\rm max}$,
    noting that the choice is not unique.
   Then we can select a complete orthonormal system $\{\ket{\omega}\}_{\omega\in \Omega}$
    as the set of all simultaneous eigenstates of the elements of $V_{\rm max}$,
    where $\Omega:=\{\omega_1,\cdots, \omega_N\}$.
   Hereafter,we identify the eigenspace of $|\omega\rangle\langle \omega |$ with $\omega$. 
   Therefore, $\Omega$ represents the way of orthogonal decomposition of the Hilbert space $\mathcal H$.
   We call the set $\{\ket{\omega}\}_{\omega\in \Omega}$ `context' induced by $V_{\rm max}$
    which is chosen from ${\cal V(N)}_{\rm max}$. 
     In due course we will use Aharonov's formalism to interpret the
      complete orthonormal system $\{\ket{\omega}\}_{\omega \in \Omega}$ as the set of the states to be `post-selected'.
  Thus, since we fix the prepared, i.e., pre-selected state to be $\ket{\psi}$ throughout this paper, 
  the `context' is the context of the measurement to be performed.
  In the case of a finite dimensional Hilbert space, a different choice of 
   $V_{\rm max}\in{\cal V(N)}_{\rm max}$ amounts to a unitary transformation of the complete orthonormal system
    $\{\ket{\omega}\}_{\omega\in \Omega}$.
    For infinite dimensional case, we have many types of von Neumann algebras ~\cite{NEUMANN,TAKESAKI}}.

We introduce the probability space $(\tilde{\Omega},\tilde{\mathcal F},\tilde{P})$ adopting the standard notation of
  Kolmogorov's measure theory of probability~\cite{Kolmo} as follows.
 The sample space is 
 $\tilde{\Omega}:=\{\omega \in \Omega \:| \: \langle \omega | \psi \rangle\neq 0 \}$, 
  and the $\sigma$-field $\tilde{\cal{F}}$ over $\tilde{\Omega}$ is the power set of $\tilde{\Omega}$.
$\tilde{P}$ is a probability measure over $\tilde{\cal{F}}$.
The measure $\tilde{P}$ satisfies the relation:
 $\tilde{P}(\{\omega_1\} \cup \{\omega_2\})=\tilde{P}(\{ \omega_1, \omega_2\})=\tilde{P}(\{\omega_1\})+\tilde{P}(\{\omega_2\})$. 
This means the probability of the event that $\omega_1$ is measured {\it or} $\omega_2$ is measured is 
the sum of $\tilde{P}(\{ \omega_1\})$ and $\tilde{P}(\{\omega_2\})$.
This probability space excludes the case
 $\{ \omega \in \Omega | \langle \omega |\psi \rangle =0\}$.
 We shall deal with the above case at the end of Sect 3.
For notational simplicity, hereafter we represent the set $\{ \omega \}\in \cal{F}$ as $\omega$ so that 
  $\tilde{P}(\omega):=\tilde{P}(\{\omega\})$.
 For a fixed choice of $V_{\rm max}$, the field ${\cal F}$ is also fixed provided that we
 specify the prepared state $\ket{\psi}$.
 
 In what follows, we use the standard notation that $\mathbb{R}$ and $\mathbb{C}$ are the fields
  of real and complex numbers, respectively, and $\sigma_x,\sigma_y,\sigma_z$ are the Pauli matrices.
\section{The main theorem}
We denote the {\it contextual value} of $A\in \mathcal{N}$ in $\omega \in \tilde{\Omega}$ by $\lambda_{\omega}(A)$
and define it as the nonzero map $\lambda_{\omega}: {\cal{N}}\to \mathbb{C}$ which satisfies the following conditions.
\begin{enumerate}
\item[(i)] Sum Rule:\\
\\
$\lambda_{\omega}(A+B)=\lambda_{\omega}(A)+\lambda_{\omega}(B), \;\;\forall A,B \in \cal{N}$\\
\item[(ii)]Product Rule:\\
\\
$\lambda_{\omega}(TS)=\lambda_{\omega}(T)\lambda_{\omega}(S),\; \;\forall T,S\in V_{\rm max}$\\
\item[(iii)]Initial condition:\\
\\
$\lambda_{\omega}(\ket{\psi}\bra{\psi})=1,\qquad \lambda_{\omega}(\ket{\psi^{\perp}}\bra{\psi^{\perp}}) = 0,$\\
\\
 where $\ket{\psi}$ is the prepared (pre-selected) state and 
$\ket{\psi^{\perp}}$ is an arbitrary state orthogonal to $\ket{\psi}$.\\

\item[(iv)]Invariance:
The expectation value and the variance of an observable $A$ defined by
\begin{eqnarray}
Ex(A):=\sum_{\omega\in \tilde{\Omega}}&\tilde{P}(\omega)\lambda_{\omega}(A),\\
Var(A):=\sum_{\omega\in \tilde{\Omega}}&\tilde{P}(\omega)|\lambda_{\omega}(A)-Ex(A)|^2
\end{eqnarray}
are independent of the choice of $V_{\rm max}$ and depends only on
the prepared state $\ket{\psi}$ and $A$.
\end{enumerate}

Some remarks on these conditions are in order. 
The sum rule (i) excludes the possibility that the values $\lambda_{\omega}(A)$ and
 $\lambda_{\omega}(B)$ are
eigenvalues of the observables $A$ and $B$, if they do not commute. 
A simple example of this is $\sigma_x$ and  $\sigma_y$ with the eigenvalues $\pm1$, while
the sum $\sigma_x+\sigma_y$ has eigenvalues $\pm \sqrt{2}$. 
Historically, the rule (i) was introduced by von Neumann~\cite{NEUMANN} when he made an argument to refute the hidden variable theory. 
The product rule (ii) means that the map $\lambda_{\omega}$ is a character on $V_{\rm max}$ and 
 implies that it can be seen as a `{\it valuation}' on $V_{\rm max}$ \cite{D}.
This means that the contextual value for an observable $T \in V_{\rm max}$ can be identified with a classical value of $T$.
We are going to generalize it to a contextual value of an observable $A\not\in {V}_{\rm max}$, 
which has no counterparts in classical theory.
 We see that $\lambda_{\omega}(\bf{1})=\rm{1}$ from (ii). 
The requirement (iii) is just an assumption based on an intuition that 
the value of the projection operator of the initial state $\ket{\psi}$ is unity and 
the values of the projection operators of the states $\ket{\psi^{\perp}}$ which are orthogonal to the prepared $\ket{\psi}$ are zero 
if we know definitely what the initial states is.
The invariance of the expectation value and the variance of an obsevable $A$ in the requirement  (iv) uniquely determines
the normal distribution function of the observable obtained after repeating many experiments according to the central limit theorem.
 
From the sum rule (i) we can apply Riez's theorem and using the normalization $\lambda_{\omega}(\bf{1})=\rm{1}$, 
we can write the contextual value of $A \in {\mathcal N}$ as
\begin{equation}
\lambda_{\omega}(A)=\frac{\Tr[W_{\omega} A]}{\Tr[W_{\omega} ]},
\end{equation}
with $W_{\omega}$ being some operator.
 The product rule (ii) for the maximal abelian subalgebra $V_{\rm max}$ narrows down the expression for the operator 
 $W_{\omega}$ to
 \begin{equation}
 W_{\omega}=a\ket{\alpha}\bra{\omega}+b\ket{\omega}\bra{\beta}+W_{\omega}^{\perp},\;\; a,b\in \mathbb{C},
\label{transverse} 
\end{equation}
for some $\ket{\alpha}\in \cal{H} $ and $\bra{\beta}\in {\cal{H^{*}}}$,
 where ${\mathcal H}^*$ is the dual space of $\mathcal H$. Here  $W_{\omega}^{\perp}$ is an operator which satisfies 
 $\{ W_{\omega}^{\perp},\ket{\omega}\bra{\omega}\}=0$.
The requirement (iii) implies that $W_{\omega}$ has a form 
$W_{\omega}=\ket{\psi}\bra{q}+\ket{r}\bra{\psi}$ for some $\bra{q}$ and $\ket{r}$.  Equating this with (\ref{transverse}),
we find that $\ket{\alpha}=\ket{\psi}$, $\bra{\beta}=\bra{\psi}$ and $W_{\omega}^{\perp}=0$ so that
\begin{equation}
 W_{\omega}=a\ket{\psi}\bra{\omega}+b\ket{\omega}\bra{\psi},\;\; a,b\in \mathbb{C},
 \label{W}
\end{equation}
and the contextual value becomes
\begin{equation}
\lambda_{\omega}(A)=\frac{a \bra{\omega}A\ket{\psi}+b \bra{\psi}A\ket{\omega}}{a \bkt{\omega}{\psi}+b \bkt{\psi}{\omega}}.
\label{value}
\end{equation}

  So far our discussion for the value of a physical quantity has not touched upon probabilistic interpretation.
  At this stage we formally introduce the concept of probability following
 the standard probability theory of  Kolmogorov~\cite{Kolmo}.
 We use the formula for the expectation value of an observable $A\in \cal{N}$,
\begin{equation}
Ex(A)=\sum_{\omega\in \tilde{\Omega}}\tilde{P}(\omega)\lambda_{\omega}(A),
\label{expectation}
\end{equation}
 where the contextual value $\lambda_{\omega}(A)$ of $A$ in $\omega\in \tilde{\Omega}$ is given by (\ref{value}).
 The expectation value
$Ex(A)$ is an average of the contextual value $\lambda_{\omega}(A)$ over the context of measurement
 $\{\ket{\omega}\}_{\omega\in \Omega}$.
The $\lambda_{\omega}(A)$ corresponds to the random variable for an observable $A$ in the probability theory.
Each $\lambda_{\omega}(A)$ depends on $\ket{\omega}$ and therefore on how we choose the context $V_{\rm max}$, but
the averaging washes out the dependence on the choice of the context,
 i.e., how we choose the set of the states to be post-selected.
  We shall show that the condition (iv) leads to a further specification of $W_{\omega}$, i.e., $b=0$ in (\ref{W}) so that
\begin{equation}
W_{\omega}=\ket{\psi}\bra{\omega}.
\label{WW}
\end{equation}

We now state and prove our main result.
\vskip 0.5cm

\begin{theo}

A map $\lambda_{\omega} : {\mathcal N} \to {\mathbb C}$ satisfying the above conditions (i)-(iv) is of the form 
\begin{equation}
\lambda_{\omega}(A)=\frac{\bra{\omega}A\ket{\psi}}{\bkt{\omega}{\psi}},\; {\rm for}\; \omega \in \tilde{\Omega},
\label{weak}
\end{equation}
and the probability measure is
 \begin{equation}
\tilde{P}(\omega)=|\bkt{\omega}{\psi}|^2.
\label{Born2}
\end{equation}
\end{theo}

\vskip 0.5cm

The contextual value (\ref{weak}) is identical to the weak value of Aharonov with $\ket{\psi}$ and $\bra{\omega}$ being the pre-selected state and the post-selected state, respectively.
By this identification, we explicitly see that the context can be interpreted as a set of post-selected states. 
Note that the probability measure $P$ is a matter of choice in the classical probability theory,
  while Born's rule (\ref{Born2}) is mandatory in quantum mechanics and even its
 non-negativity is an important aspect of the consequence. The asymmetry of
  $\bra{\omega}$ and $\ket{\psi}$ in ({\ref{WW}}) comes from the expression for
 the expectation value ({\ref{expectation}) in which the post-selected states are summed up on
  the basis of an intuition that we can choose a pre-selected state but
 its outcome is uncertain. The arrow of time is built-in in (\ref{expectation}).
 The expression (\ref{expectation}) for the expectation value with
(\ref{WW}) and (\ref{Born2}) reproduces the standard quantum mechanical expectation value,
\begin{equation}
Ex(A)=\bra{\psi}A\ket{\psi},
\label{expectationvalue}
\end{equation}
which certainly does not depend on the context $\{\ket{\omega}\}_{\omega\in \Omega}$.
 The apparent context dependence of $Ex(A)$ in (\ref{expectation}) goes away  by
the completeness relation on $\ket{\psi}$: $ \sum_{\omega\in \tilde{\Omega}}\langle \psi \ket{\omega}\bra{\omega} A |\psi \rangle
 = \bra{\psi}A\ket{\psi}$.
It is instructive to think about the possibility 
$\tilde{P}(\omega)=|\bkt{\omega}{\psi}|^4$ instead of (\ref{Born2}). Then we would have
$Ex(A)=\sum_{\omega\in {\tilde{\Omega}}}|\bkt{\omega}{\psi}|
^4\frac{\bra{\omega}A\ket{\psi}}{\bkt{\omega}{\psi}}=\sum_{\omega\in {\tilde{\Omega}}}|
\bkt{\omega}{\psi}|^2\bkt{\psi}{\omega}
\bra{\omega}A\ket{\psi}$. The context dependence would not go away. We can follow  the exact parallel way also for
the variance $Var(A)$ to confirm its context independence. 
\vskip 0.5cm

\begin{proof}
\vskip 0.5cm
Consider a functional
\begin{equation}
{\cal L}_{E}(\bra{\omega},\mu):=Ex(A)-\sum_{i=0}^{N-1}\mu_{i}[\sum_{\omega\in {\tilde{\Omega}}}\bkt{\psi_{i}}{\omega}
\bra{\omega}A\ket{\psi }-\bra{\psi_{i}}A\ket{\psi}], \qquad \qquad \mu_i \in \mathbb{R}, 
\label{lagrangian}
\end{equation}
where $\mu_{i}$ are the Lagrange multipliers and $\{\ket{\psi_{i}}\}$ is a complete orthonormal system which contains $\ket{\psi }=\ket{\psi_{0}}$ as a unit vector.
The  Lagrange multiplier terms give constraints
\begin{equation}
\sum_{\omega\in {\tilde{\Omega}}}\bkt{\psi_{i}}{\omega}
\bra{\omega}A\ket{\psi }-\bra{\psi_{i}}A\ket{\psi}=0,\; \;\forall \;i,\;\; \forall A\in \cal{N}.
\label{constraints}
\end{equation}
Since $\{\ket{\psi_{i}}\}$ is a  complete orthonormal system and $A\in \cal{N}$ is arbitrary, the above constraints 
are equivalent to the completeness relation $\sum_{\omega\in \Omega}\ket{\omega}\bra{\omega}=1$ for $\{\ket{\omega}\}_{\omega\in \Omega}$ as we see comparing with
its matrix elements with (\ref{constraints}) .

We demand that the variation of ${\cal L}_E(\bra{\omega},\mu)$ with respect to $\bra{\omega}$ should vanish. That is,
\begin{eqnarray*}
\frac{\delta{\cal L}_E(\bra{\omega},\mu)}{\delta \bra{\omega}}=&\frac{\partial \tilde{P}(\omega)}
	{\partial \bra{\omega}}\frac{a \bra{\omega}A\ket{\psi}+b \bra{\psi}A\ket{\omega}}{a \bkt{\omega}{\psi}
		+b \bkt{\psi}{\omega}}+\tilde{P}(\omega)\frac{a A\ket{\psi}}{a \bkt{\omega}{\psi}+b \bkt{\psi}{\omega}}\\
			&\;\;\;\;-\;\;\tilde{P}(\omega)\frac{a \bra{\omega}A\ket{\psi}+b \bra{\psi}A\ket{\omega}}
                        {(a \bkt{\omega}{\psi}+b \bkt{\psi}{\omega})^2} a\ket{\psi}-\sum_{i}\mu_{i}\bkt{\psi_{i}}{\omega}A\ket{\psi}=0
.\;\;\; \forall A\in \cal{N}
\end{eqnarray*}

A particular choice $A=\ket{\psi}\bra{\psi}$ implies that $\frac{\partial \tilde{P}(\omega)}{\partial \bra{\omega}}
\propto \ket{\psi}$, since the second, third and fourth terms are then $\propto \ket{\psi}$. To see the implication
of the component of the above equation  perpendicular to $\ket{\psi}$ consider the case $A=\ket{\psi^{\perp}}\bra{\psi}+\ket{\psi}\bra{\psi^{\perp}}$,
 where $\ket{\psi^{\perp}}$ is a state orthogonal to $\ket{\psi}$ to
find that the second and the forth terms cancel each other.
 We therefore see that the sum of the first and the third terms and the sum of
  the second and the fourth terms separately vanish, i.e.,
\begin{eqnarray}
&\frac{\partial \tilde{P}(\omega)}{\partial \bra{\omega}}-\frac{\tilde{P}(\omega)}{a \bkt{\omega}{\psi}+b \bkt{\psi}{\omega}} a\ket{\psi}=0,\\
&\tilde{P}(\omega)\frac{a}{a \bkt{\omega}{\psi}+b \bkt{\psi}{\omega}}-\sum_{i}\mu_{i}\bkt{\psi_{i}}{\omega}=0.
\label{second}
\end{eqnarray}
Adding these two equations yields
\begin{equation}
\frac{\partial \tilde{P}(\omega)}{\partial \bra{\omega}}=\sum_{i}\mu_{i}\bkt{\psi_{i}}{\omega}\ket{\psi}
\end{equation}
and its integration gives $\tilde{P}(\omega)=\sum_{i}\mu_{i}\bkt{\psi_{i}}{\omega} \bkt{\omega}{\psi}+\tilde{P}_0$, where $\tilde{P}_0$
 is an integration constant. Therefore, (\ref{second}) becomes
\begin{equation}
(\sum_{i}\mu_{i}\bkt{\psi_{i}}{\omega} \bkt{\omega}{\psi}+\tilde{P}_0)\frac{a}{a \bkt{\omega}{\psi}+b \bkt{\psi}{\omega}}
-\sum_{i}\mu_{i} \bkt{\psi_{i}}{\omega}=0, \forall {\omega}\in \tilde{\Omega}.
\end{equation}
 This implies that $\tilde{P}_0=0$ and $b=0$. We note that $\frac{\delta{\cal L}_E(\bra{\omega},\mu)}{\delta \ket{\omega}}=0$ is automatically satisfied.
 Therefore, the context invariance of the expectation value $Ex(A)$ leads to expressions for the probability measure and the contextual value of an observable $A$ as
 \begin{eqnarray}
&\tilde{P}(\omega)=\sum_{i}\mu_{i}\bkt{\psi_{i}}{\omega} \bkt{\omega}{\psi},\\
&\lambda_{\omega}(A)=\frac{\bra{\omega}A\ket{\psi}}{\bkt{\omega}{\psi}}.
\label{exp}
\end{eqnarray}
 We further demand that the variance $Var(A)=\sum_{\omega\in \tilde{\Omega}}\tilde{P}(\omega)|\lambda_{\omega}(A)|^2$ is context independent.
 To simplify the discussion we set $Ex(A)=0$ or replace $A$ by $A-Ex(A)$. 
 It is possible to set up a similar variational problem for $Var(A)$ to the previous one for $Ex(A)$. Instead we simply insert the result of (\ref{exp}) into
 the expression for $Var(A)$  to obtain
 \begin{eqnarray}
Var(A)&=\sum_{\omega\in \tilde{\Omega}}\tilde{P}(\omega)|\lambda_{\omega}(A)|^2\nonumber \\
&=\sum_{\omega\in \tilde{\Omega}}\sum_{i}\mu_{i}\bkt{\psi_{i}}{\omega} \bkt{\omega}{\psi}|\frac{\bra{\omega}A\ket{\psi}}{\bkt{\omega}{\psi}}|^2 \nonumber \\
&=\sum_{\omega\in \tilde{\Omega}}\sum_{i}\mu_{i}\frac{\bkt{\psi_{i}}{\omega}}{{\bkt{\psi}{\omega}} }\bra{\omega}A\ket{\psi}\bra{\psi}A\ket{\omega}.
\end{eqnarray}
This is independent of the context $\{\ket{\omega}\}_{\omega\in \Omega}$ if and only if $\mu_{i}=0$ for $i\neq 0$. 
The remaining parameter $\mu_{0}$ is fixed to be unity by the normalization condition $\sum_{\omega\in \tilde{\Omega}}\tilde{P}(\omega)=\mu_{0}=1$.
 Then we arrive at the main theorem (\ref{weak}) and (\ref{Born2}).
  \end{proof}

Now we can naturally extend the probability space $(\tilde{\Omega},\tilde{\mathcal F},\tilde{P})$ to 
$(\Omega, {\mathcal{F}, P})$ as follows.
 The sample space is  ${\Omega}$, and the $\sigma$-field $\mathcal F$ over $\Omega$ is the power set of $\Omega$.
${P}$ is the probability measure over ${\cal{F}}$ satisfying the following conditions: 
$P|_{\tilde{\mathcal{F}}}=\tilde{P}, P|_{{\mathcal F}\setminus  \tilde{\mathcal F}}=0$.
From the theorem, we can write $P(\omega)=|\langle \omega | \psi \rangle|^2$.
 Thus the probability space $(\Omega, \mathcal{F}, P)$ describes the 
 standard quantum mechanical statistics in the context $\{\ket{\omega}\}_{\omega\in \Omega}$.

 \section{Summary and Discussions}
    We have shown that the contextual value of an observable is the weak value by demanding the consistency of quantum mechanics
with  Kolmogorov's measure theory of probability in conjunction with the consideration of the values of obsevables. 
The crucial requirement was that  the expectation value  and the variance of an observable should be independent of the way of measurement.
This leads eventually  to Born's rule in quantum mechanics. The assumptions of the sum rule (i) and the product rule (ii) may not have much problem and
the invariance (iv) of the expectation value and the variance is crucial to relate the contextual value to the concept of probability. The third assumption (iii) is debatable. At the moment we do not have
complete justification. We believe the usefulness of (iii) to characterize the contextual value. 

    Going back to the original motivation in the introduction, we would like to examine in what sense the weak value can be regarded as the value before projective measurements.
    Let $A(t)$ be an operator at time $t\leq T$ in the Heisenberg representation, where $T$ is the time of the projective measurement. We choose $A(T)$ as an element of $V_{\rm max}$
    so that the post-selected state $\bra{a}$ is an eigenstate of $A(T)$, $\bra{a}A(T)=\bra{a}a$ of the eigenvalue $a$. 
  The weak value
\begin{equation}
\frac{\bra{a}A(t)\ket{\psi}}{\bkt{a}{\psi}}\; 
\label{before}
\end{equation}
can be interpreted as the value before projective measurement provided that $\bra{a}$ is not orthogonal to the pre-selected state $\ket{\psi}$.
At $t=T$, the weak value (\ref{before}) coincides with the eigenvalue $a$.
As stressed before the weak value depends on the context, i.e., the states to be post-selected. 
The price is that the contextual value is complex in general.

 The extension of the derivation of Born's rule to the mixed state case is straightforward but technically involved
  so that we postpone it for future publication.

  From the main theorem we see that the value of an observable is given by the weak value (\ref{weak}).
    As is by now well known, the weak value is
  experimentally accessible ~\cite{RSH,PCS,RLS,Pryde,WSZLHG,RESCH,HK,YYKI,PMNBVEK,BS} and theoretically analyzed by many people~\cite{MJP,JOZSA, SH,HS}. 
  The weak value coincides with one of eigenvalues when the observable is restricted to the maximal abelian subalgebra $V_{\rm max}\in{\cal V(N)}_{\rm max}$.
  For a general $A\in \cal{N}$, the weak value can be deduced from the information associated with the state to be post-selected, which is circumstantially inferred from the set of eigenvalues 
  to be obtained by projective measurements of all the elements of $V_{\rm max}$.

The Kochen-Specker theorem~\cite{KS} tells us that it is not possible to assign non-contextual values to all observables in quantum mechanics.
 Since the only non-contextual values are eigenvalues of the observables, 
 this theorem does not apply to the weak values. Actually the weak value is explicitly dependent on the
 context defined by the post-selected states.

 In the proof of Gleason's theorem~\cite{Gleason}, he demanded that the measure over
  the set of all the subspaces of the Hilbert space ${\cal H}$ should be independent of the choice of basis
 in order to conclude that the measure has a quadratic form in the state vector components if dim$({\cal H})\geq 3$. 
 This leads to the standard expression for the expectation value. There is a gap
 between the probability and the measure  over
  the set of all the subspaces of the Hilbert space ${\cal H}$, so that the derived Born's rule lacks the probabilistic interpretation. In a mathematical sense, Gleason's theorem is close to ours.  
  We believe that we have filled the gap by introducing the physical  concept of the contextual value of observables defined by the prepared state and the states to be post-selected. 
  A rather surprising finding in the present work is that the contextual value coincides with the weak value.

   Zurek~\cite{Zurek} claimed that he derived Born's rule from ``the environment assisted invariance".
    He considered an entangled state 
\begin{equation}
\ket{\psi}=\frac{\ket{s_1}\ket{e_1}+\ket{s_2}\ket{e_2}}{\sqrt{2}},
\end{equation}
where $\ket{s_1}$ and  $\ket{s_2}$ are the orthonormal basis of the system, while 
 $\ket{e_1}$ and $\ket{e_2}$ are the orthonormal basis of the environment.
Consider an observable: $A\otimes {\bf 1}$ with $A=\ket{s_1}\bra{s_1}-\ket{s_2}\bra{s_2}$. 
 The elementary events are ${x_1}:=\{\bra{s_1}\bra{e_1}\}$ and ${x_2}:=\{\bra{s_2}\bra{e_2}\}$.
 The sample space is $\Omega':=\{ x_1,x_2\}$, 
 and the $\sigma$-field is the power set of $\Omega'$.
 We would like to show that the probabilities to obtain ${x_1}$ and ${x_2}$ are $1/2$. 
 To see this we compute the weak values:
 $\lambda_{x_1}(A)=1,\;\lambda_{x_2}(A)=-1$. We apply the formula (\ref{expectation}) noting that
  the left hand side is zero because of symmetry under the SWAP operation of
 the states of 1 and 2 both for the system and environment.
  The operator $A\otimes {\bf 1}$ does not distinguish the state of environment so that the expectation value
 has to be symmetric under the SWAP of the system state.
  Then we see that $P({x}_1)-P({x}_2)=0$ on the right had side and arrive at the equal probability $P({x_1})=P({x_2})=1/2$,
    using $Ex({\bf1})=P({x_1})+P({x_2})=1$.
 In this simple case, only the discrete symmetry is sufficient to derive Born's rule.
  Here we see that the symmetry principle should apply to the expectation value to obtain
 Born's rule. The role of entanglement with the environment is not clear for us,
  though we suspect a possible connection to the post-selection.

\section*{Acknowledgments}
The present work has been inspired by the stimulating lecture  by Professor Andreas D{\"o}ring at Nagoya.
The authors would like to acknowledge the useful comments of Professor Masanao Ozawa and information on the types of 
von Neumann algebras by Professor Izumi Ojima and deeply thank Professor Shin Takagi for his critical comment which leads to
a substantial improvement of the previous version.
The authors are supported by the Global Center of Excellence Program ``Nanoscience and Quantum Physics" at Tokyo Institute of Technology.


\end{document}